# New view on the diffraction discovered by Grimaldi and Gaussian beams

**A. Yurkin**

M. Prokhorov General Physics Institute of RAS,

e-mail: alvl1yurkin@rambler.ru

In offered work short historical excursus to the classical theory of light is presented: Grimaldi, Fermat, Newton, Huygens, Young, Fresnel, Fraunhofer, and Gauss. The ray analog of wave model of light and Huygens-Fresnel's elementary waves on the basis of consideration of geometrical model is offered. New geometrical properties of Gaussian beams are analyzed. The new, generalized interpretation of a corner of diffraction divergence of beams of light is given. Difference of geometrical properties of wave fronts of infinite and finite length is shown. Examples of possible application of our geometrical model in various areas are given.

## Introduction

In work [24] the descriptive geometry optic model on the basis of consideration of binomial distribution for the description of distribution of light in the laser was offered. Calculations were carried out when using paraxial (Gaussian) approach. Actually in work the descriptive geometrical figure representing from a position of the theory of graphs [9] "tree"[1] which number of branches on each (conditionally chosen) step doubles was offered [24]. Such approach corresponds to the term "diffraction" (from Latin "diffringere" – to break, split in two ("di-ffraction")) introduced by the Italian physicist Grimaldi (1665) who has discovered this light phenomenon. Grimaldi inclined to a wave explanation of the nature of light [7, 17, 22].

[1] Our "tree" has some geometrical particularities: very long trunk (symmetry axis) with the branches inclined under small angles to a trunk and remote on small distances (in comparison with length of branches) from a trunk. In a further statement we don't have need to use the theory of graphs as we deal with the concrete geometrical figure consisting of direct pieces and corners.

In work [12] the general description of a nonlinear arithmetic pyramid and arithmetic triangles of the second (nonlinear) type was given and recurrent formulas for creation of a nonlinear arithmetic pyramid and nonlinear arithmetic triangles were offered, examples from combination theory are given.

In works [13, 15] the binomial of the second, non-Newtonian type was offered, new communication of binary notation and binomial coefficients of different types was shown. Geometrical interpretation of binomials of two types by consideration of trajectories of rays for a two-dimensional case is given.

Numerical calculations of binomial distribution of the second type are given in work [14] for big degrees of a binomial. The analogy to binomial distribution of the first type is drawn. In work [16] analogy between systems of geometrical and arithmetic triangles of different types was shown.

In the offered work short historical excursus to the classical theory of light is presented at the time of Grimaldi and the subsequent development of the theory connected with names Fermat, Newton, Huygens, Young, Fresnel, Fraunhofer, and Gauss. The ray analog of wave model of light and Huygens-Fresnel's elementary waves on the basis of consideration of geometrical model is offered. New geometrical properties of Gaussian beams are analyzed. The new, generalized interpretation of a corner of diffraction divergence of light is given. Difference of geometrical properties of wave fronts of infinite and final lengths is shown.

Let's note as essential simplification in work on this article existence of data containing in [23].

Let's note also as very useful by the preparation of this work studying of the latest theory of V. S. Leonov stated in [18] and key of a problem of modern physics, stated in "Leonov's list" [19], supplementing "Ginzburg's list" [3]. However these publications don't contain a mention of possibility of research of the phenomenon of diffraction throughout Grimaldi's ideas (as, for example, in work [7]) and don't indicate the need of further research of geometrical properties of the structures leading to Gaussian or normal distribution [5], and also Gaussian beams [6]. In this regard we will quote work [8, page 49]: "The central limit theorem and its generalizations explain why in the nature normal distribution meets so often, especially, in connection with sizes which are made of many ("almost") equally distributed ("almost") independent random components. It is necessary to emphasize that in the nature such "compositions" aren't always formed of random variables by their sum, therefore studying of behavior of other functions very important. Poincare somehow noticed with sarcasm that all believe in universality of normal distribution: physicists trust because think that mathematics proved its logical need, and mathematicians trusts as consider that physics checked it by laboratory experiments".

Gaussian beams, including having normal distribution, are widely applied in optics, including in laser equipment [4, 6]. In the present work we investigate new geometrical models (structures) of the Gaussian beams leading to normal distribution.

V. L. Ginzburg with a great interest treated my works on a related topic - light distribution in the lasers, reported at his seminar in 1998. I am grateful to him for his remarks and the help.

## 1. Short excursus to the classical theory of light at the time of Grimaldi and the subsequent

### 1.1. Fermat, Newton

The geometrical optics is the simplest idea of light nature; by means of its approaches it is possible to solve a set of problems of lighting engineering and optoequipment [6]. The light beam in geometrical optics is abstract mathematical concept, and the geometrical optics is the limit case of wave optics corresponding vanished to small length $\lambda$ of a light wave $\lambda \to 0$.

The geometrical optics is based on the principle of the shortest optical way (or the minimum time of distribution). It was formulated by Fermat as the general law of distribution of light (1660). For a homogeneous medium this principle is reduced to the law of rectilinear distribution of light.

Newton also adhered to the theory of rectilinear distribution of light in the form of the light particles flying rectilinearly. [6] For an explanation of the phenomenon of diffraction Newton entered other term - "inflection", i.e. a curvature of beams.

**1.2 . Huygens**

The main part of the light phenomena is explained by the wave theory based on the *principle of Huygens* (1690). The wave theory is based on the law of rectilinear distribution of light [6]. However this law loses force or has additional restrictions of applicability at the description of such phenomena as diffraction of light [6].

The simplified illustration of the principle of Huygens is given in Fig. 1:

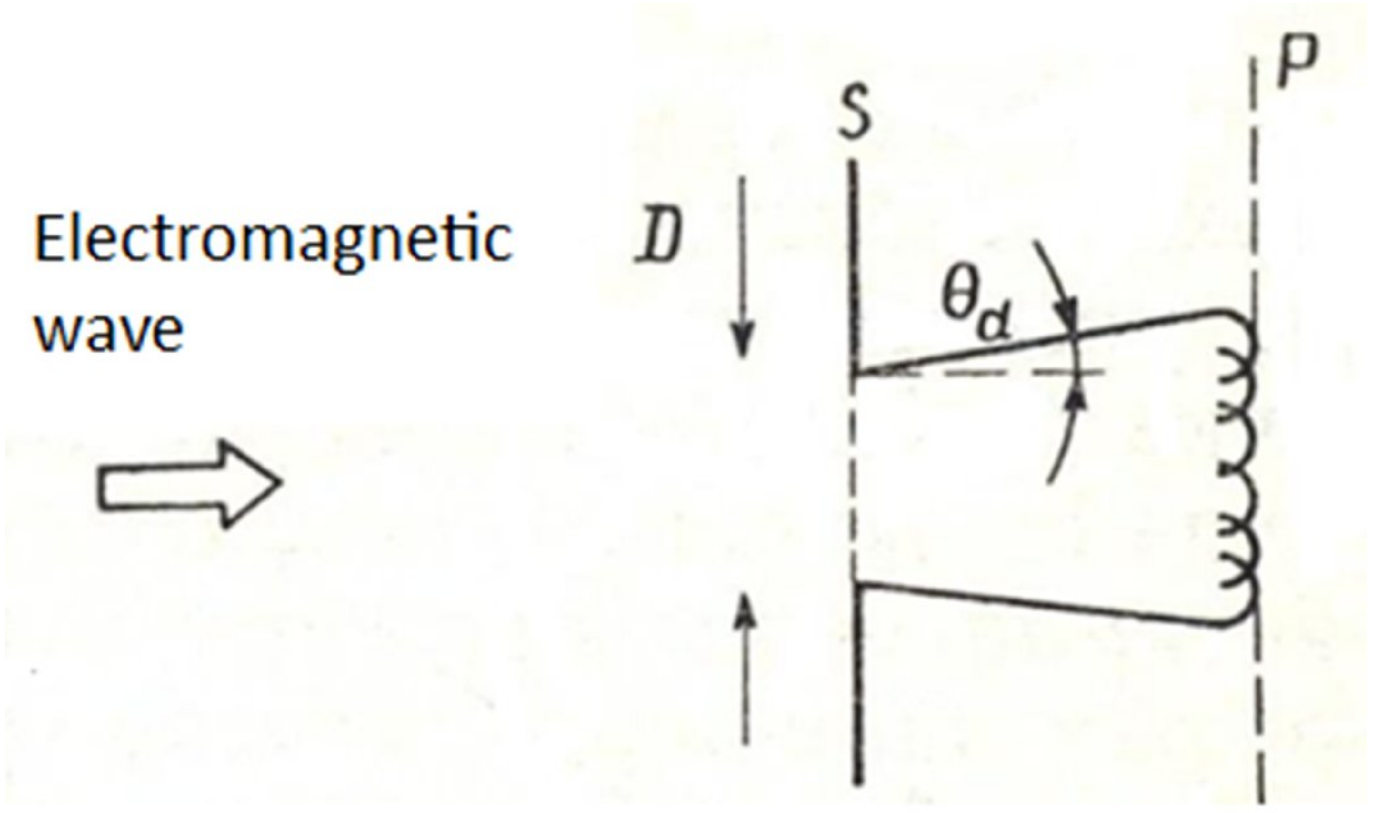

Fig. 1. Illustration of the principle of Huygens and creation of the flat wave front (Fig. 1.6. from the monograph [4]).

In this drawing the beam with constant intensity and the flat wave front falls on the screen $S$ in which there is an opening with a diameter of $D$. "According to Huygens's principle the wave front in some plane $P$ behind the screen can be received by superposition of the elementary waves radiated by each point of an opening. We see that because of the final size $D$ of an opening the beam has final divergence $\theta_d$. It can be calculated by means of the diffraction theory"[4, page 21]: $\theta_d \approx \frac{1}{2}\lambda/D$. Let's note that in practice [24] it is more convenient to apply expression:

$$\theta_D \approx \lambda/D, \tag{1}$$

where $\theta_D = 2\theta_d$.

Huygens speaks about distribution of the wave front in the form of a geometrical surface i.e. when length of a wave is infinitely small in comparison with the extent of the wave front. Therefore Huygens's principle is in essence the principle of geometrical optics, and light extends in the form of spherical or flat surfaces [6]. *It is obvious that points of radiation of elementary waves are located on the line of the wave front at some distance of* $b$ *from each other.*

### 1.3. Young

Historically the first wave treatment of diffraction was given by Young (1800). Except the law of distribution of the wave front in the direction of beams, Young were entered by the principle of transfer or diffusion of amplitude of fluctuations along the wave front, i.e. across beams. Speed of this transfer of diffusion on Young is proportional to length of a wave and grows with increase in distinction of amplitudes in the next points of the wave front. Thus, in process of distribution of the wave front there is a smoothing of non-uniform distribution of amplitude on the wave front, and diffusion of amplitude is accompanied by change of a phase of fluctuations. The method of Young was developed further in detail in the diffusive theory of diffraction [6, 7, 22].

### 1.4. Fresnel

*Fresnel's hypothesis* (1815) consists in an assumption of dependence of amplitude of secondary waves from an angle $\varphi$ between a normal of $n$ to an auxiliary surface of $S$ and the direction on a point of observation of $B$. In Fig. 2 the drawing explaining this assumption is provided, decrease of amplitude is presented by reduction of thickness of a curve.

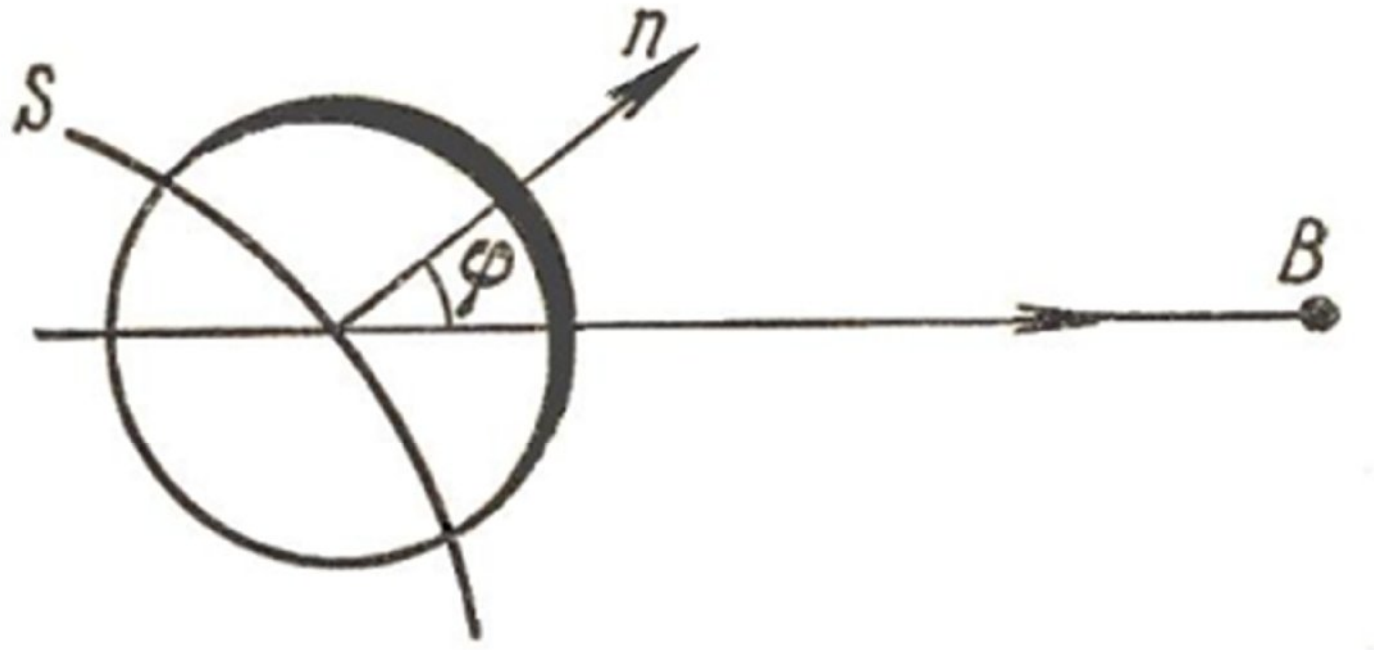

Fig. 2. Illustration of a hypothesis of Fresnel (Fig. 8.21. from the monograph [6]).

According to this assumption, amplitude decreases in process of angle increase $\varphi$ and becomes equal to zero when absolute size is equal or more $\pi/2$.

Fresnel investigated the phenomena of diffraction of spherical or flat waves in the point of observation laying at final distance from an obstacle therefore such diffraction phenomena usually name *Fresnel's diffraction* [6].

### 1.5. Fraunhofer

Fraunhofer investigated a bit different type of diffraction namely in parallel beams (1822). This type of diffraction usually name Fraunhofer's diffraction [6]. Essentially Fraunhofer's diffraction doesn't differ from Fresnel's diffraction, however consideration of a case of diffraction of Fraunhofer mathematically isn't difficult, but it is important at research of the questions concerning operation of optical devices.

### 1.6. Gauss

**1.6.1.** Gauss who much has made for progress in development of optical tools, ignored the wave theory: "The most important lack of work of Gauss was never taken by it into account in general: Gauss approached to all optical problems from positions of the naive corpuscular theory of light and didn't show any interest to the theory of distribution of light of Fraunhofer" [2, page 149]. "Gauss made some remarks on "the wave theory" of light, but he never dealt with this issue. It must be kept in mind that the Gauss was familiar with Fraunhofer since the trip to Munich in 1816" [2, page 199].

**1.6.2.** Diffraction angle $\theta_d$ (Fig. 1) for many practical calculations of beams of light can consider small, i.e. $\sin\theta_d \approx \theta_d$ therefore the front $P$ in Fig. 1 can be considered approximately flat, and a beam - *paraxial* or *Gaussian*. In Gaussian optics "points and the rays lying in close proximity to an axis are considered; the members containing squares and higher degrees of distances from an axis or corners between beams and an axis, are rejected" [1, page 186].

**1.6.3.** In Fig. 3 the Gaussian beam is presented:

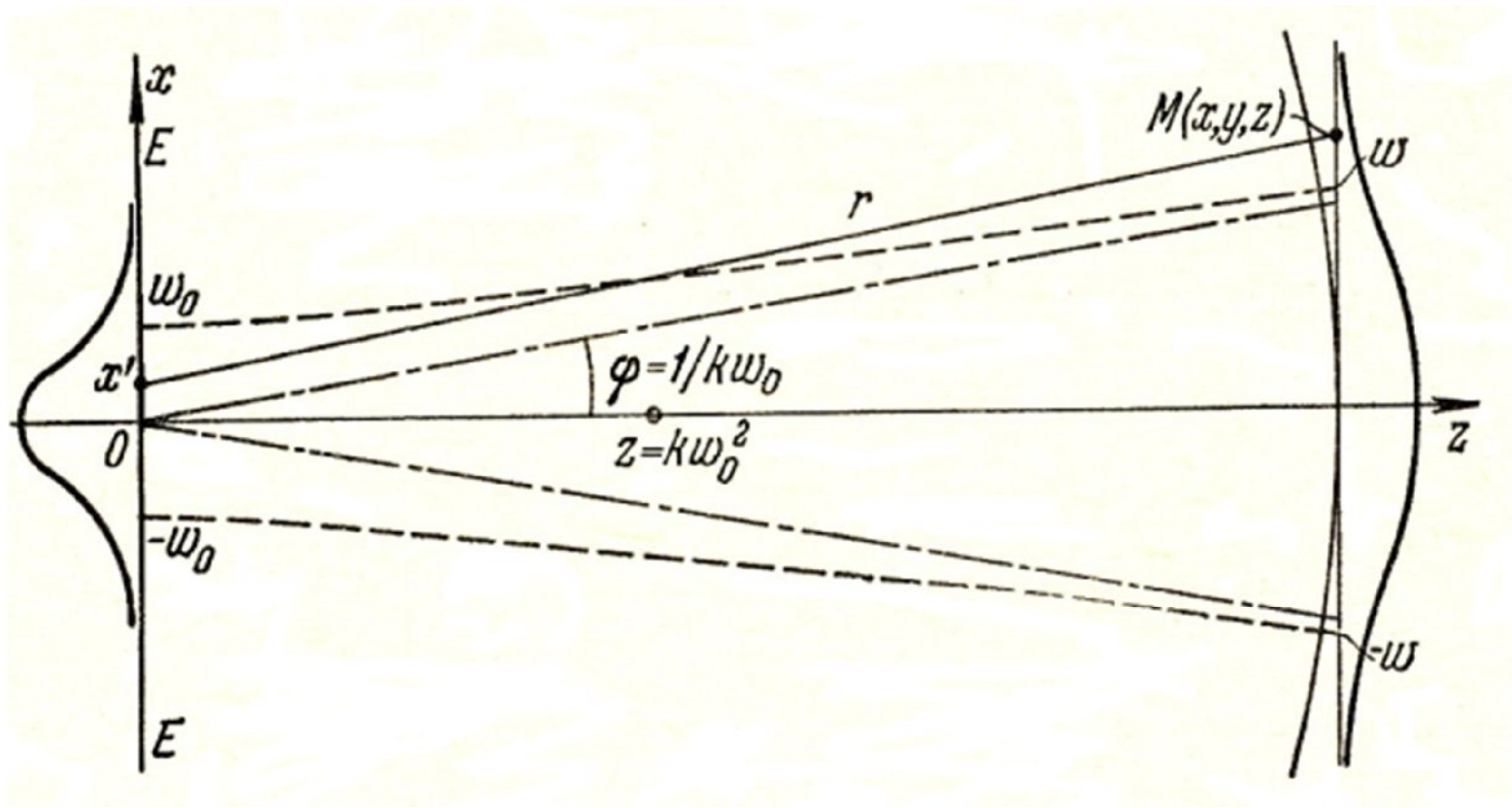

Fig. 3. Gaussian beam: distribution of amplitude of fluctuations of the wave front is described by Gauss function. The $EE$ plane is a surface of the wave front, (Fig. 9.8. from the monograph [6]).

At the left and on the right in Fig. 3 schedules of distribution of amplitude are presented to the planes of the wave front described by function of Gauss:

$$a(x',y') = a_0 \exp(-\frac{x'^2+y'^2}{2w_0^2}). \quad (2)$$

$M(x,y,z)$ is an observation point with coordinates $x,y,z$ and $r$ is distance from a point $(x', y', 0)$ to a point of $M(x,y,z)$.

The size $w_0$ defines area of change of $x',y'$, where intensity of fluctuations is proportional, $a^2(x', y')$, decreases in $e$ of times in comparison with the maximum value, $a^2$ reached at $x' = 0, y' = 0$. Thus, $w_0$ characterizes area in which energy in the $EE$ [6] plane is concentrated.

Gaussian beams have a number of features.

1. The diffraction picture is characterized by monotonous reduction of intensity and doesn't contain oscillations and lines of zero intensity (diffraction strips), characteristic for diffractions on openings.

2. If distribution of amplitude of a field looks like a Gaussian curve, the radius of curvature of the wave front and width of distribution of amplitude, instead of a curve form changes only.

3. The example of a Gaussian beam can serve as fine interpretation of diffraction as diffusions of amplitude (item 1.3) of a field along the wave front in process of its distribution in the medium.

## 2. Geometry of flat wave fronts

In our researches we operate with two main concepts: 1) the wave (light) front is a straight line; 2) ray is a straight line, perpendicular to the line of the wave front, i.e. a normal to a front line.

Thus, offered geometrical models of distribution of wave (light) fronts consist only of straight lines, points of intersection of lines and angles between these lines.

### 2.1. Infinitely extended front

The image of flat infinitely long wave front $F$ is given in Fig. 4:

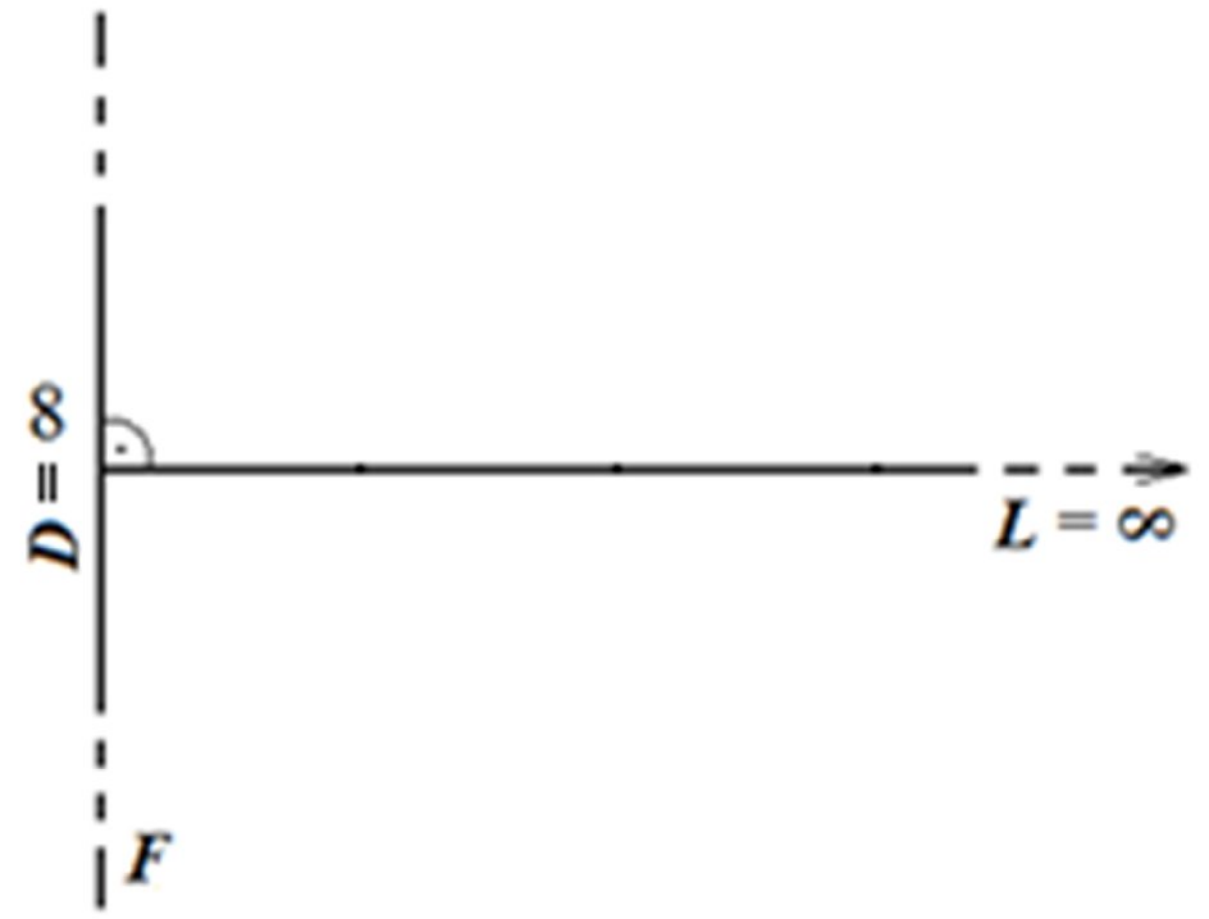

Fig. 4. Infinitely long flat wave front.

The front $F$ extends along a ray of $L$. Length of the wave front $D = \infty$, length of a ray of $L = \infty$, the ray (normal) is perpendicular the front plane. Space we consider uniform. As $D = \infty$, divergence (broadening) of a light beam is absent regardless of the size of length $\lambda$ of a wave.

### 2.2. Front of finite length

**2.2.1.** In offered model we consider paraxial (Gaussian) approach to research of light beams (item 1.6).

If the wave (light) front $F$ has the finite length, that is $D = d$, it is possible to make an assumption (postulate) that the ray (normal) along which extends the front, too has the finite length of $l$. Within $l$ distance such model can describe approximately process of distribution of the flat front (Fig. 5):

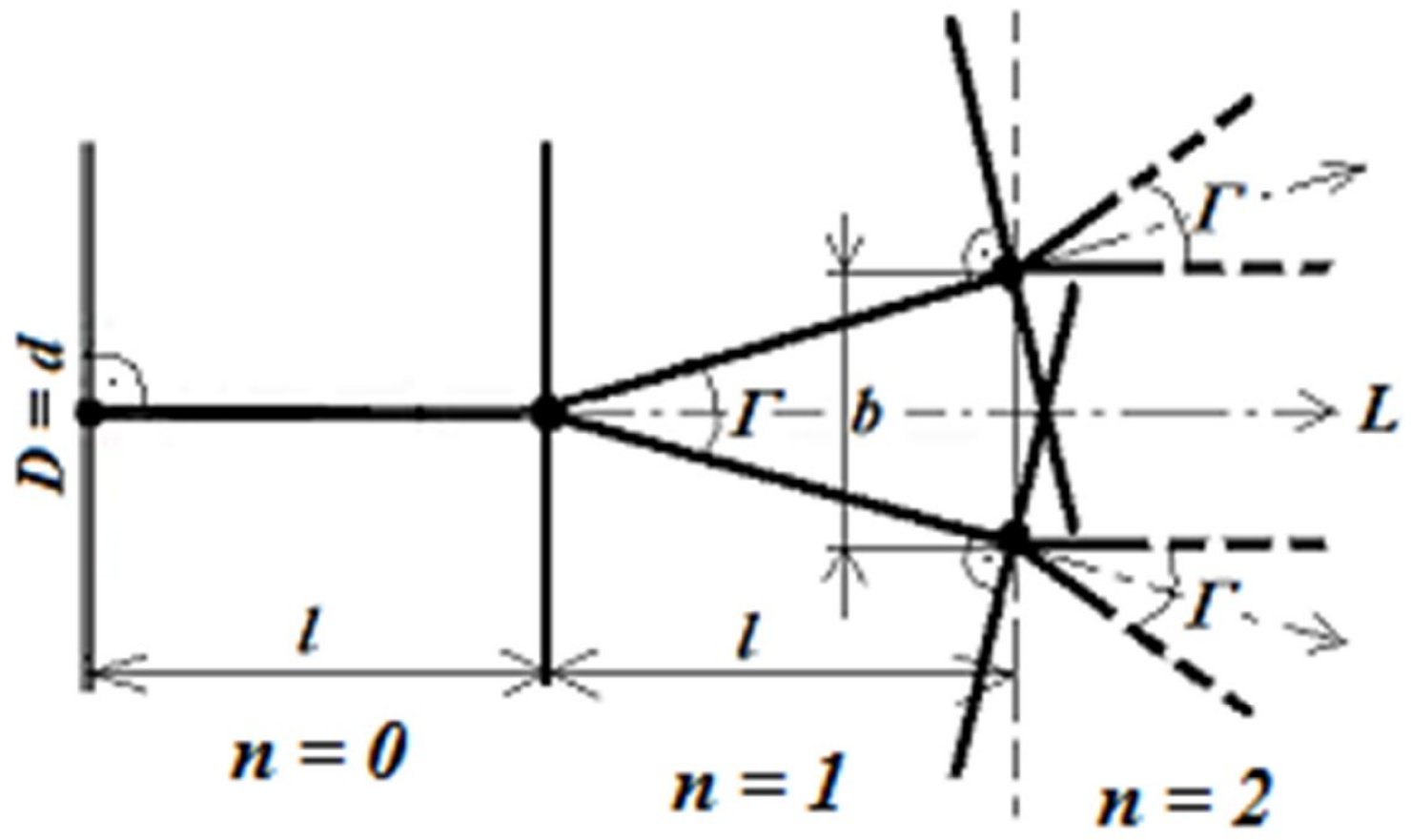

Fig. 5. Flat wave front of finite length.

Having passed $l$ distance for some time of $t$, we consider (in our model) that the ray is split symmetrically in two, instead of one ray, we have two rays with a small angle Γ between them. In Fig. 5 these two rays, at everyone the front are shown. After pass of the following distance of $l$ for some time of $t$ beams again are symmetrically split in two etc. Generally it is possible to accept $l \approx L/n$.

Numbers of passes of rays in Fig. 5 are designated by $n$, $n = 0, 1, 2$ .... Initial pass (without splitting) we will consider by zero ($n = 0$). Vertical dashed line in Fig. 5, we will call this line the *conditional front* (the similar question was investigated in [21]), shows an approximate arrangement of fronts after $n = 1$ pass. *Distance between points of branching (points of intersection) of lays with this vertical line equally to* $b$, (by analogy to item 1.2). As the angle Γ is small, $l \gg b$, $d \gg b$.

In works [11, 12, 24] we investigated the branching systems of rays consisting of groups of rays, connected in the broken trajectories consisting of links.

In Fig. 6 the group of branching rays of $K$ inclined on angles of $\pm\frac{1}{2}k\Gamma$ to the axial line is shown; an angle $\Gamma = 4\,\gamma$, $k = 0, 1, 2$ ....

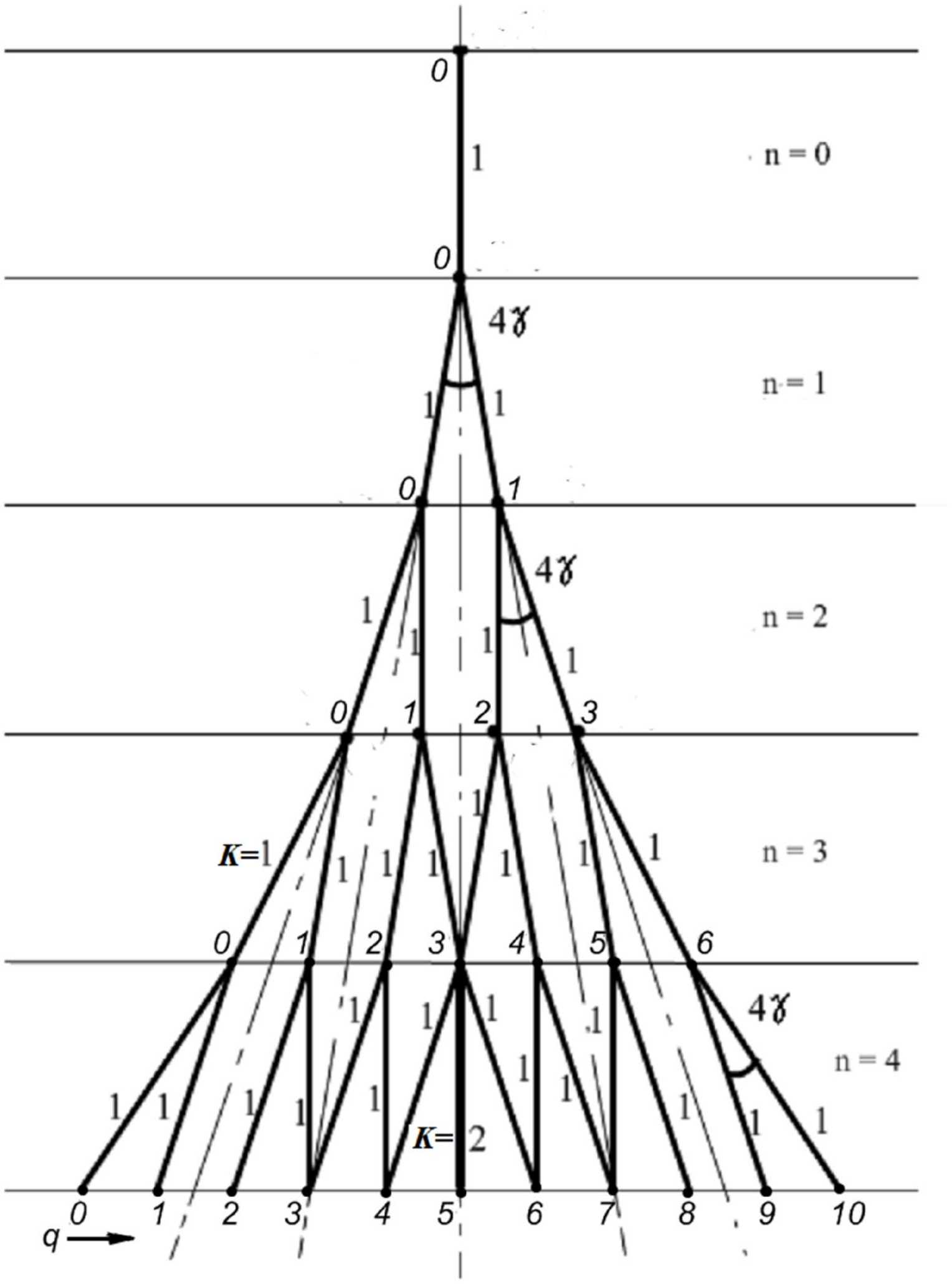

Fig. 6. System of the rays inclined under angles of $\pm\frac{1}{2}k\Gamma = \pm 2k\gamma$ to a vertical. Rays of $K$ and links of $N$ are shown by continuous lines, symmetry axes by the dash-dotted lines. Light extends from top to down, $n$ – number of pass of rays, $q$ – number of points of branching of rays. The distance between the next points of branching is equal to $b$.

After each pass of light of $n$, (Fig. 6) rays branch in two when crossing the line of the conditional front. Distribution of the light energy leaving along rays, is proportional to quantity of rays. The first three passes energy is distributed between rays equally. Part of rays is parallel each other, i.e. are inclined on identical angles. Rays of $K$ extend along $N$ links.

Since the fourth pass ($n = 4$), some rays are imposed at each other, energy distribution between rays becomes nonuniform. Generally the number of rays of $K$ is more or equally to number of links of $N$, i.e. $K \geq N$. In Fig. 6 the number of rays of $K$ is marked with numbers, and points of $q$ of branching of rays are marked in the italics.

In Fig. 7 the example of distribution of the rays branching in a point of $q = 3$ between $n = 3$ and $n = 4$ is presented:

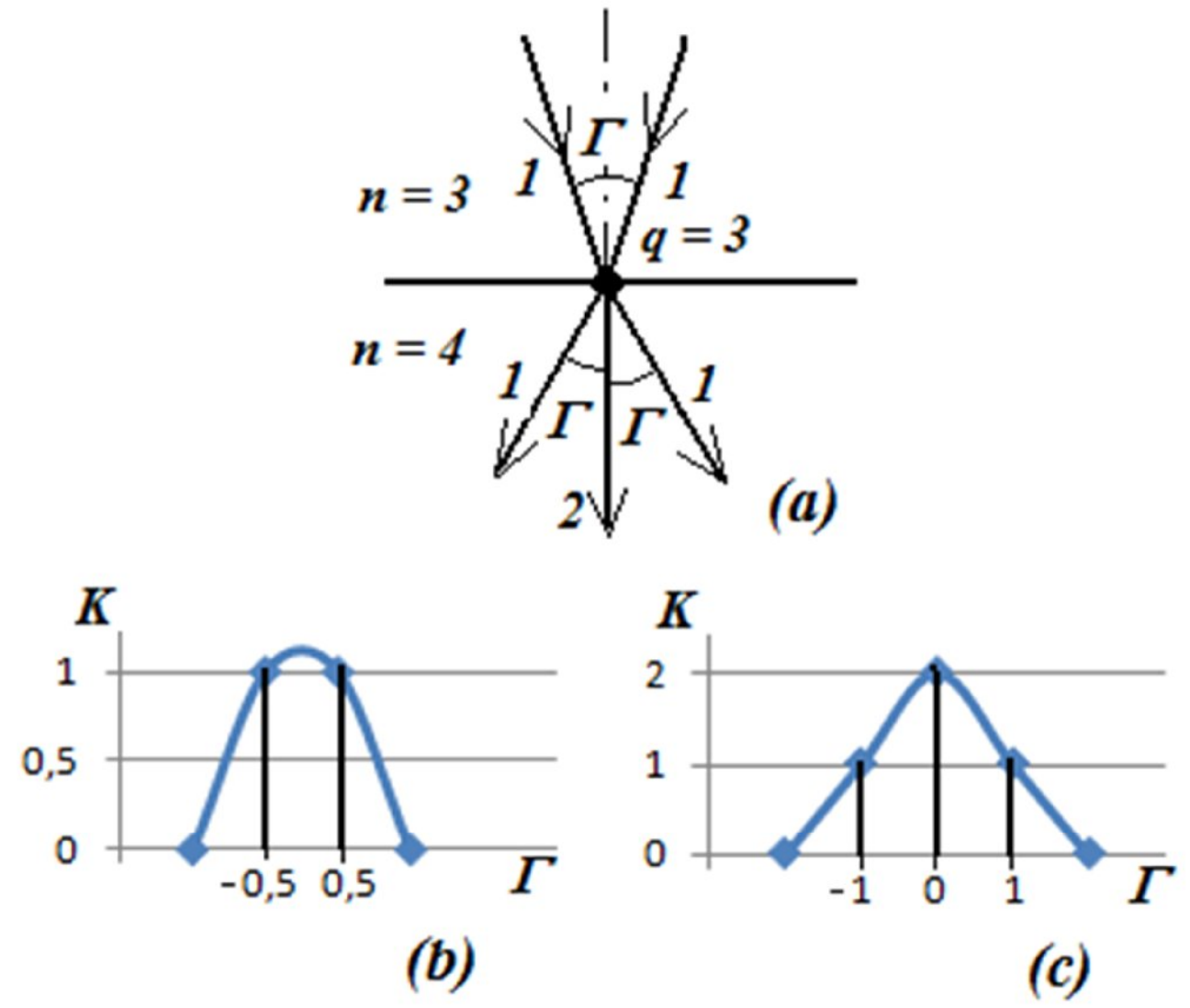

Fig. 7. The rays entering and leaving a point of branching of $q = 3$ (a), angular distribution of entering (b) and leaving (c) of rays.

In Fig. 7a rays branching entering and leaving a point (the same point is noted in Fig. 6 between $n = 3$ and $n = 4$ passes in the drawing center) are shown. Angular distribution of number of the rays entering into a point of branching in Fig. 7b leaving in Fig. 7c.

At increase in number of passes of $n$ the number of rays entering and leaving branching points, increases, and a form of rays envelope around of angular distribution (the energy extending along rays) tends to the normal.

**2.2.2**. In work [12] the system of rays represented in Fig. 6 was considered, and calculations of construction three-dimensional and the two-dimensionals numerical tables showing the general distribution of all rays within pass of $n$ were carried out.

In the present work we will give new calculations of distribution of entering and leaving rays in points of their branching.

**2.2.3. The Rays entering into points of branching of $q$.** Layers (two-dimensional tables) *the nonlinear arithmetic pyramid of the first type* offered in [12] are presented in Fig. 8:

$\rightarrow q$ ; $p\downarrow$ ; $\binom{n}{p}$ ; $\Gamma$

**[n=0]**

| $p \backslash q$ | 0 | $\binom{n}{p}$ | $\Gamma$ |
|---|---|---|---|
| 0 | 1 | 1 | 0 |

**[n=1]**

| $p \backslash q$ | 0 | 1 | $\binom{n}{p}$ | $\Gamma$ |
|---|---|---|---|---|
| 0 | 1 | 0 | 1 | 0,5 |
| 1 | 0 | 1 | 1 | –0,5 |

**[n=2]**

| $p \backslash q$ | 0 | 1 | 2 | 3 | $\binom{n}{p}$ | $\Gamma$ |
|---|---|---|---|---|---|---|
| 0 | 1 | 0 | 0 | 0 | 1 | 1 |
| 1 | 0 | 1 | 1 | 0 | 2 | 0 |
| 2 | 0 | 0 | 0 | 1 | 1 | –1 |

**[n=3]**

| $p \backslash q$ | 0 | 1 | 2 | 3 | 4 | 5 | 6 | $\binom{n}{p}$ | $\Gamma$ |
|---|---|---|---|---|---|---|---|---|---|
| 0 | 1 | 0 | 0 | 0 | 0 | 0 | 0 | 1 | 1,5 |
| 1 | 0 | 1 | 1 | 1 | 0 | 0 | 0 | 3 | 0,5 |
| 2 | 0 | 0 | 0 | 1 | 1 | 1 | 0 | 3 | –0,5 |
| 3 | 0 | 0 | 0 | 0 | 0 | 0 | 1 | 1 | –1,5 |
| $\begin{bmatrix} n \\ q \end{bmatrix} \rightarrow$ | 1 | 1 | 1 | 2 | 1 | 1 | 1 | | |

**[n=4]**

| $p \backslash q$ | 0 | 1 | 2 | 3 | 4 | 5 | 6 | 7 | 8 | 9 | 10 | $\binom{n}{p}$ | $\Gamma$ |
|---|---|---|---|---|---|---|---|---|---|---|---|---|---|
| 0 | 1 | 0 | 0 | 0 | 0 | 0 | 0 | 0 | 0 | 0 | 0 | 1 | 2 |
| 1 | 0 | 1 | 1 | 1 | 1 | 0 | 0 | 0 | 0 | 0 | 0 | 4 | 1 |
| 2 | 0 | 0 | 0 | 1 | 1 | 2 | 1 | 1 | 0 | 0 | 0 | 6 | 0 |
| 3 | 0 | 0 | 0 | 0 | 0 | 0 | 1 | 1 | 1 | 1 | 0 | 4 | –1 |
| 4 | 0 | 0 | 0 | 0 | 0 | 0 | 0 | 0 | 0 | 0 | 1 | 1 | –2 |

Fig. 8. Distribution of the rays entering into points of branching of $q$. Layers of a nonlinear arithmetic pyramid for various values $n$. $p$ are numbers of rows, and $q$ are number of columns for each of layers. In the first column outside rectangles on the right – the sums of numbers of the rows $p$, expressed usual binomial to coefficients $\binom{n}{p}$, on the right are given in the second column the value of tilt angles of rays (Fig. 6), multiple to angles $\pm\Gamma/2$. For $[n = 3]$ the sums of numbers in the vertical columns, expressed in binomial coefficients of a type $\begin{bmatrix} n \\ p \end{bmatrix}$ [12] are shown. The number sequence, inclined on 45° (are highlighted in bold type) corresponds to distribution of the rays leaving points of branching [12].

Numerically angular distribution of the beams entering into points of branching of $q$, is shown in the form of the numbers making columns in rectangles in Fig. 8.

**Example 1**. Entering under corners $\pm\Gamma/2$ in a point of branching of $q = 3$ of rows given in Fig. 7a, b it is possible to see angular distribution in Fig. 8 – a column for $[n = 3]$ – numbers $0, 1, 1, 0$ are located in a vertical frame.

The rule of consecutive filling with numbers of our tables (Fig. 8) – pyramid layers, since top:

$$\begin{bmatrix} n \\ p \\ q \end{bmatrix} = \begin{bmatrix} n-1 \\ p-1 \\ q-p \end{bmatrix} + \begin{bmatrix} n-1 \\ p \\ q-p \end{bmatrix}. \qquad (3)$$

In detail process of filling of these tables (with examples) is shown in [12].

Angular distribution entering into points of branching of rays (Fig. 6) at $n$ pass, i.e. number of rays under angles, multiple $\pm\Gamma/2$ is equal to the sum of numbers of a row $p$ of a layer of $n$ (in Fig. 8 the numbers to the right of a frame) is expressed by usual binomial coefficients $\binom{n}{p}$.

**2.2.4. The rays leaving points of branching of *q*.** Layers (two-dimensional tables) *nonlinear arithmetic pyramid of the second type are* presented in Fig. 9:

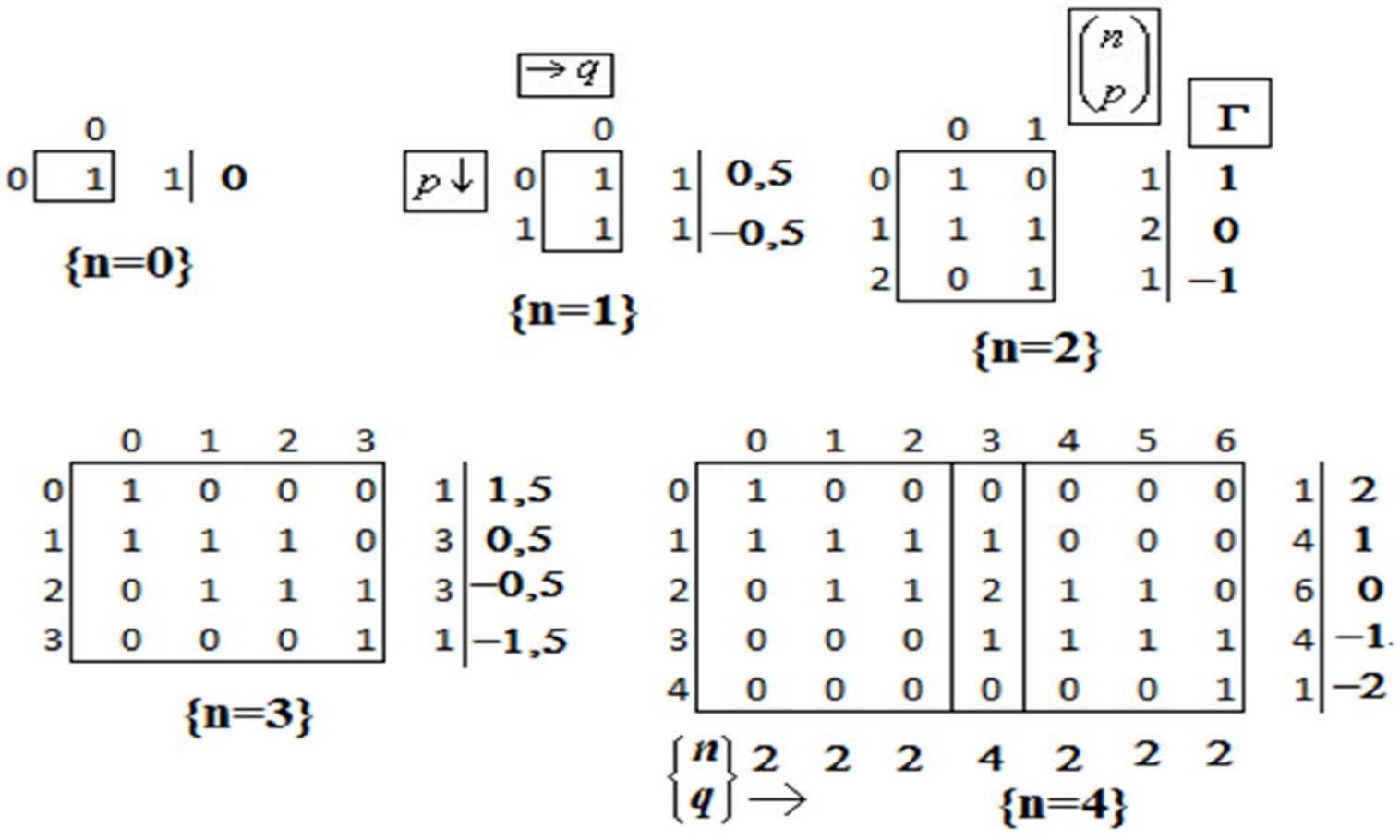

Fig. 9. Distribution of the rays leaving the points of branching of $q$. Layers of a nonlinear arithmetic pyramid for various values $n$. $p$ – numbers of rows, and $q$ – number of columns for each of layers. In the first column outside rectangles on the right – the sums of numbers of the rows $p$, expressed usual binomial to coefficients $\binom{n}{p}$, on the right are given in the second column of value of tilt angles of rays (Fig. 6), multiple to angles $\pm\Gamma/2$. For $\{n = 4\}$ the sums of numbers in the vertical columns, expressed in binomial coefficients of a type $\begin{Bmatrix} n \\ p \end{Bmatrix}$ [12] are shown.

Numerically angular distribution of the rows leaving points of branching of $q$, is shown in the form of the numbers making columns in rectangles in Fig. 9.

**Example 2.** Angular distribution of the rows leaving under angles $\pm\Gamma/2$ a point of branching of $q = 3$ of rows (it is given on Fig. of 7 a, c) it is possible to see in Fig. 9 – a column for $\{n = 4\}$ – numbers $0, 1, 2, 1, 0$ are located in a vertical frame.

The rule of consecutive filling with numbers of our tables (Fig. 9) – pyramid layers, since top, by analogy to expression (3):

$$\begin{Bmatrix} n \\ p \\ q \end{Bmatrix} = \begin{Bmatrix} n-1 \\ p-1 \\ q-p+1 \end{Bmatrix} + \begin{Bmatrix} n-1 \\ p \\ q-p \end{Bmatrix}. \tag{4}$$

Angular distribution in leaving points of branching of rows (Fig. 6) at $n$ pass, i.e. number of rays under angles, multiple $\pm\Gamma/2$ is equal to the sum of numbers of a row $p$ of a layer of $n$ (in Fig. 9 the numbers to the right of a frame), is expressed by usual binomial coefficients $\binom{n}{p}$.

**3. About similarity and distinction of models describing light propagation**

**3.1.** According to item 1.1: in our ray researches we used constructions only by means of straight lines and corners but didn't use such concepts as curves, circles areas, etc.

Such approach which has been actually based on the way to consideration of diffraction as splitting, the offered Grimaldi in our model (Fig. 5-7), assumes that rays can repeatedly be split and not always extend rectilinearly or on the shortest distance.

**3.2.** In item 1.2 it was noted that Huygens describes propagation of the wave front by means of superposition of the elementary wave fronts proceeding from points. In our rays researches we used concepts of the paraxial rays entering and proceeding from points of their branching (Fig. 5-7). And we assume that process of branching of rays take place in our model, in that case if the wave front has the finite length (Fig. 5).

**3.3.** In item 1.3 it was noted Young entered by the principle of diffusion of amplitude of fluctuations along the wave front, i.e. across beams. From our geometrical constructions (Fig. 6) process of "diffusion" of [7, 20, 22] light rays along the wave front in process of light distribution is also visually visible.

**3.4.** In item 1.4 the hypothesis of Fresnel explaining elementary secondary waves of Huygens's by means of circles, made of the curves having the unequal thickness (Fig. 2) was described. Our researches showed that it is possible to use the elementary geometrical objects – straight lines entering and proceeding from branching points for the similar description of process of distribution of light (Fig. 5-7).

**3.5.** According to item 1.5: Fraunhofer's researches belong to studying of properties of parallel beams. Our geometrical constructions too investigate properties of actually parallel beams since we use paraxial or Gaussian approach.

**3.6.** According to item 1.6: Consideration of our descriptive geometrical model with binomial coefficients of $\begin{bmatrix} n \\ q \end{bmatrix}$ and $\begin{Bmatrix} n \\ q \end{Bmatrix}$ types [12, 14], and also research of distributions of beams entering and leaving points (Fig. 7), illustrates variety of the geometrical models leading to Gaussian (normal) distribution.

**3.7.** In work [10] for the characteristic of laser "quasiresonator" [11], the dimensionless geometrical (mechanical) number of $m = D/4\gamma L$ by analogy to Fresnel's wave number $F = 4D^2/L\lambda$ [4] was offered. Where: $D$ is laser aperture, i.e. width of the wave front, angle $\gamma = \Gamma/4$ and $L$ is length of the laser resonator. It is possible to find correspondence between $m$ and $F$ numbers if we accept $\Gamma = 4\gamma \sim \lambda/D$, and then $m \sim F/4 = D^2/\lambda L$. Therefore it is possible to find correspondence between two couples orthogonal pieces (Fig. 5): wave front long $D$ and length of a ray of $l$, and also wave length (extending along rays) $\lambda$ and $b$ distance between points of branching of rays. The generalized expression (1) then we will write down as:

$$\theta_D \approx \frac{\lambda}{D} \approx \frac{b}{l} \approx \Gamma. \tag{5}$$

**Conclusion**

In the present work we showed the general regularities and gave evident descriptive interpretation of wave and ray theories of light and Huygens-Fresnel's elementary waves, considering geometrical properties of only one figure represented in Fig. 6. This figure can be three-dimensional [21], however in its basis only thin lines, dimensionless points and small angles lie. Our simple and evident model works in paraxial (Gaussian) approach is suitable for the description of features of distribution of narrowly directed Gaussian beams of laser.

This model leading to Gaussian distribution can be used for evident geometrical interpretation of distribution of light in homogeneous mediums, and mediums containing scattering centers [11]. This model also can be used and for interpretation movement of bodies in the mediums containing small, evenly distributed heterogeneities approximately of the identical size. For example for the description of possible trajectories of movement of the arrow which has been let out from bow during a rain, trajectories of movement of the drill when drilling deep wells, etc.

It is possible to refer description, simplicity of calculations, system communication with known and new [5, 12, 13] mathematical ratios to advantages of the offered geometrical model (normal distribution, binomial formulas, binary notation, arithmetic triangles of various types, etc.), historical continuity with Grimaldi's ideas who for the first time has explained diffraction as process of splitting of light. Unfortunately, our model, as well as any other, can't apply for the description of all variety of light and other physical phenomena, however promotes their understanding in all depth and variety.

**Acknowlegements**

The author expresses his gratitude to prof. E.E.Shnol and to prof. V. G. Mikhalevich for reading of the manuscript and for the useful remarks stated by preparation of this work.